\documentclass{article}

\usepackage{PRIMEarxiv}

\usepackage[utf8]{inputenc} 
\usepackage[T1]{fontenc}    
\usepackage{hyperref}       
\usepackage{url}            
\usepackage{booktabs}       
\usepackage{amsfonts}       
\usepackage{nicefrac}       
\usepackage{microtype}      
\usepackage{lipsum}
\usepackage{fancyhdr}       
\usepackage{graphicx,bm,algorithmic,subcaption,textcomp,amsmath,amssymb,amsfonts,caption,setspace}       
\graphicspath{{media/}}     

\title{GHM Wavelet Transform for Deep Image
Super Resolution

}

\author{
  Ben Lowe, Hadi Salman, Justin Zhan \\
  Computer Science Department \\
  University of Arkansas \\
  Fayetteville, Arkansas\\
  \texttt{\{bjlowe, hs028, jzhan\}@uark.edu} \\}

\begin{document}
\maketitle

\begin{abstract}
The GHM multi-level discrete wavelet transform is proposed as preprocessing for image super resolution with convolutional neural networks. Previous works perform analysis with the Haar wavelet only. In this work, 37 single-level wavelets are experimentally analyzed from Haar, Daubechies, Biorthogonal, Reverse Biorthogonal, Coiflets, and Symlets wavelet families. All single-level wavelets report similar results indicating that the convolutional neural network is invariant to choice of wavelet in a single-level filter approach. However, the GHM multi-level wavelet achieves higher quality reconstructions than the single-level wavelets. Three large data sets are used for the experiments: DIV2K, a dataset of textures, and a dataset of satellite images. The approximate high resolution images are compared using seven objective error measurements. A convolutional neural network based approach using wavelet transformed images has good results in the literature. 
\end{abstract}


\section{Introduction}
Image super resolution is the task of creating a high resolution image from a low resolution input. There exist several applications that require super resolution such as medical and satellite image reconstruction and video recording applications \cite{parksurvey}. What makes image super resolution challenging is the lack of information to completely characterize a high resolution output. As such, there are many choices for interpolating pixels with trade-offs in computational complexity and restoration quality. Super resolution techniques can take multiple low resolution images or a single low resolution image as input \cite{vansurvey}. This paper is focused on investigating the properties of wavelets in single image super resolution achieved with convolutional neural networks. 

There are three main problems with super resolution techniques: blurring of sharp edges, blocking artifacts in diagonal edges, and limited high frequency reconstruction for fine detail \cite{vansurvey}. Simple interpolation techniques, such as bicubic interpolation, which, while popular methods, are not considered true super resolution techniques because they cannot reconstruct high frequency data in the image signal which constitutes fine detail in the image \cite{parksurvey}. Several learning based approaches have been proposed in the literature with excellent reconstruction quality \cite{dwsr, vdsr, srcnn, learningbaseddwt, liu2019multi, wvDWDN, wvBlindDeblur, wvSRNet, wvNN}.

This paper investigates the effects of 37 different single-level wavelet transformations of images used as filtering for convolutional neural networks to achieve single image super resolution. The convolutional neural networks predict the difference between the coefficients of the wavelet decomposition for the input image and the high-resolution ground truth. This paper proposes the use of the GHM multi-level wavelet as a superior image representation to achieve deep image super resolution. The GHM wavelet outperforms single-level wavelet representations, while the single-level wavelet representations behave similarly. This indicates that the structure of the representation is more important than how well the filter responses maintain properties of the image important to the human-visual system such as edge preservation and texture preservation.

The rest of this paper is structured as follows. Section \ref{sec:relatedworks} discusses previous work in super resolution with a focus on wavelet based techniques. Section \ref{sec:methods} outlines the mathematical background, the architecture of the convolutional neural network used and a brief overview of image quality assessments used in this experiment. Section \ref{sec:experiment} presents seven objective error measurements between the 37 single-level wavelets from the Haar, Daubechies, Biorthogonal, Reverse Biorthogonal, Coiflets, and Symlets wavelet families as well as the GHM multi-level wavelet. Section \ref{sec:conclusion} concludes this work with recommendations for future work.

\section{Related works}
\label{sec:relatedworks}
\subsection{Brief Taxonomy of Super Resolution Techniques}
There are many different super resolution techniques which can be broadly categorized into Multiple Image Super Resolution (MISR) and Single Image Super Resolution (SISR) \cite{nassurvey}. MISR uses multiple similar images as input for the reconstruction of one or more high resolution output images while SISR uses a single image as input for reconstruction \cite{nassurvey}. Some techniques for multiple image super resolution include iterative back projection\cite{cohen2000polyphase}, iterative adaptive filtering\cite{farsiu2006practical}, direct methods\cite{farsiu2004fast}, projection onto convex sets\cite{akgun2005super}, maximum likelihood\cite{jung2011position}, and maximum a posteriori methods\cite{zhang2012generative}. Single image super resolution techniques include learning-based approaches and reconstruction approaches which include primal sketches\cite{sun2003image}, gradient profile\cite{yan2015single}, wavelet\cite{learningbaseddwt}, contourlet transform\cite{yang2012self}, bilateral filters\cite{dai2007bilateral}, and Gaussian mixture models\cite{ogawa2012super}. This paper focuses on single image super resolution by learning coefficients for wavelet approximation.

\subsection{Wavelet Based Single Image Super Resolution}
The use of wavelets on images has been thoroughly researched in literature. One of the main uses of wavelets is to enhance an image by extracting the features, enhancing it and reconstructing the image to a higher resolution version. Karthikeyan et al. \cite{wvUS} proposed an image enhancement system that utilized wavelets as follows: denouncing with wavelet shrinkage, edge enhancement with Sigmoid function, and image fusion using Bicubic interpolation. Zhou \cite{wv2D} proposed a 2D wavelet reconstruction algorithm for image enhancement. The original image is duplicated and scaled then 2D wavelet decomposition is applied. Using Discrete Fourier Transform instead of bicubic interpolation was proposed for the high frequency information reconstruction.

Another use of wavelets is for image denoising to maintain the features of an image and reducing the energy of the noise. Ye et al. \cite{wvDenoise} used wavelet coefficients to represent images which can be optimized to reduce the influence of noise. Wavelets was also used for optimized compression of images. Cheng et al. \cite{wvDecomp} proposed an adaptive wavelet transform via image texture. Chappalli et al. \cite{wv2Gen} studied second generation wavelet super resolution through a trade-off between the blur introduced and noise in reconstructed images. Celik et al. \cite{wvDualTree} proposed a dual-tree complex wavelet transform to enhance both geometric and intensity features of imagery.

Demirel et al. \cite{wvDisStat} proposed multi-level wavelet transform using discrete and stationary decomposition. The low resolution image is processed twice each with different decomposition method yielding 4 sub-bands each. The duplicate sub-bands are then accumulated and inverse transformed to the high resolution image. Chavez-Roman et al. \cite{wvEdge} proposed a resolution-enhancement technique that augment the fine features extracted from the wavelet's high frequency sub-band using sparse interpolation.

Single Image super resolution often uses learning to supplement the lack of information contained in the input image. Learning based algorithms using wavelet transforms attempt to learn the coefficients for a high resolution approximation of an input low resolution image \cite{dwsr}.

An approach by Jiji et al. in 2004 learned wavelet coefficients by constructing a database of coefficients for patches of filtered and transformed images \cite{jiji2004single}. The process iterates over each $4 \times 4$ block and chooses a known coefficient block with minimum absolute distance from the training database \cite{jiji2004single}. Jiji et al. uses a three level transform using the Daubechies 4 wavelet for the image processing \cite{jiji2004single}. The choice of the optimal wavelet is not investigated in their work.

Another approach by Lui, Wu, Mao, and Lien in 2007 learns wavelet coefficients choosing the best matching coefficients from a database of projection weights constructed from the training set. A Markov network with the maximum a posteriori estimation approach is used to select the best matching coefficients for each patch of the high frequency sub-bands of the transform\cite{lui2007learning}. Lui et al. uses a two level transform using the Haar wavelet for the image processing and does not consider the use of other wavelets\cite{lui2007learning}.

A technique to linearly interpolate wavelet coefficients was proposed by Tong and Leung in 2007 for image super resolution. Tong uses a two dimensional Taylor series expansion to approximate the coefficients of each wavelet subband by computing the weighted sum of Taylor series coefficients from a set of low resolution images $\{I\prime_1,I\prime_2,\ldots,I\prime_m\}$ \cite{tong2007super}. The set of low resolution images are constructed by appyling spatial shifted operators on the low resolution input image \cite{tong2007super}. Tong uses a single level Haar wavelet transform but their method can utilize any wavelet, although they do not investigate their properties. Tong \cite{tong2007super} proposes if the relation between the high and low resolution images is given by the function $f$, then the reconstruction is given by
\begin{equation}
\centering
I_{l_1l_2}(i,j) = \sum_{k_2=l_2}^{w-1+l_2}\sum_{k_1=l_1}^{h-1+l_1}f(\bm{x}_R + s_{k_1k_2})/(hw). \notag
\end{equation}

\subsection{Image Super Resolution with Neural Networks}
Zhang et al. \cite{wvNN} proposed a deep learning method based on discrete wavelet transform to reduce noise in audio records. As a conclusion, Zhang et al. promoted the use of different wavelets for deep learning and the success of wavelet based learning in noise reduction. 
A wavelet based deep neural network for super resolution was proposed by Tiantong Guo et al \cite{dwsr}. Using the wavelet sub-bands as input, the proposed technique learns the residuals of the wavelet transform to fill in the missing details. Results developed in their work outperform other non wavelet based learning approaches for image super resolution. The paper encourages further examination of different wavelets for super resolution.    
In \cite{wvSRNet}, Huang et al. developed wavelet based convolution neural network for Face super resolution. The experiment starts by extracting the wavelet coefficients and apply a prediction task in deep learning framework. Three loss functions were used: wavelet prediction loss, texture loss, full-image loss. The technique achieved promising results for Gaussian blur, pose and occlusions.  
Min et al. developed a method that combines both wavelet transform and deep convolution network to reduce image deblurring. Min et al. recommended further exploration of different wavelets and their effect on image restoration. 
As an effort to apply learning based wavelet denoising, Li et al. \cite{wvDWDN} was able to utilize the wavelet frequency characteristics to lower the noise in medical images.
Recent work in 2019 by Lui et al. proposes using a multi-level discrete wavelet transform convolutional neural network, called MWCNN, and applies it to super resolution, image denoising, and JPEG artifact removal with results similar to the state-of-the-art \cite{liu2019multi}. Lui et al. only consider the Haar wavelet \cite{liu2019multi} and the wavelet performance analysis presented in this work could further improve the performance of MWCNN.
\section{Methodology}
\label{sec:methods}
This section gives a brief overview of the mathematics of the wavelet transform as well as the objective error measurements used to analyze the performance of the analyzing wavelets. Then a discussion of the subjective measurements is given with examples of each artifact. The network architecture and hyperparameters of the neural network is presented.
\subsection{Single-Level Wavelets}
\label{singlewavelets}
\begin{figure}
	\centering
	\includegraphics[width=0.7\columnwidth]{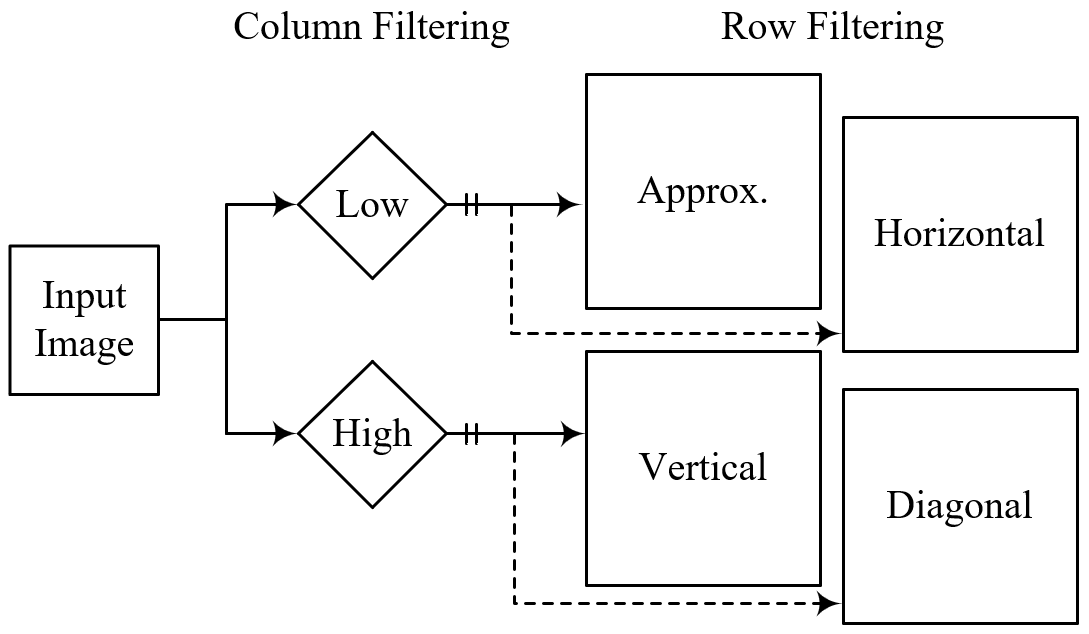}
	\caption{A 2-D wavelet decomposition with four output bands.}
	\label{fig:dec}
\end{figure}

\begin{figure}
	\centering
	\includegraphics[width=0.7\columnwidth]{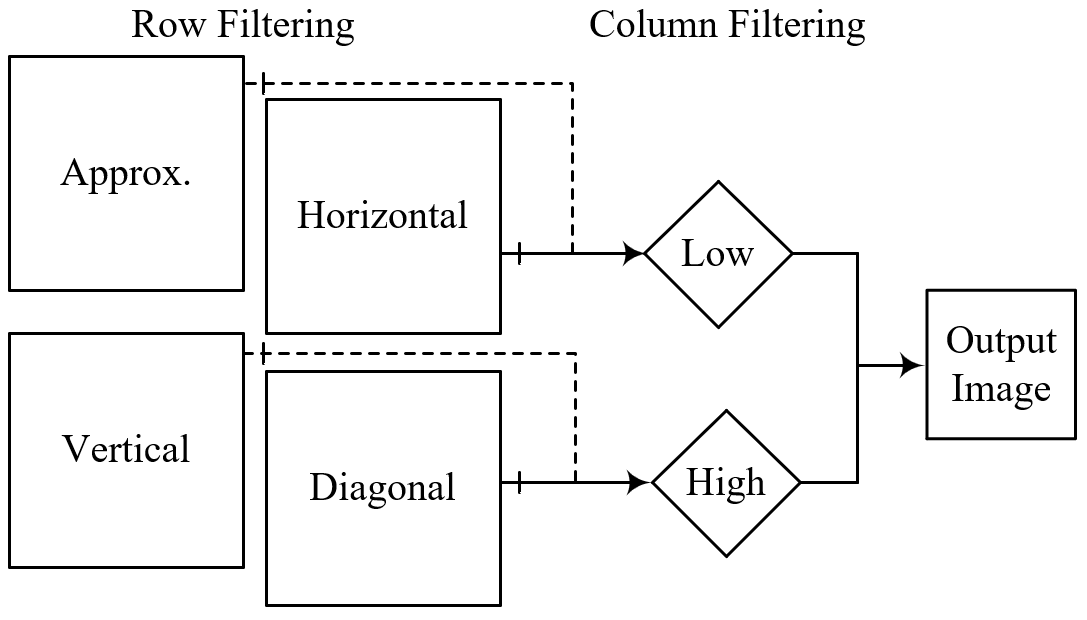}
	\caption{A 2-D wavelet reconstruction combining four input signals.}
	\label{fig:rec}
\end{figure}

\begin{figure}
	\centering
	\begin{subfigure}{0.4\columnwidth}
		\includegraphics[width=\columnwidth]{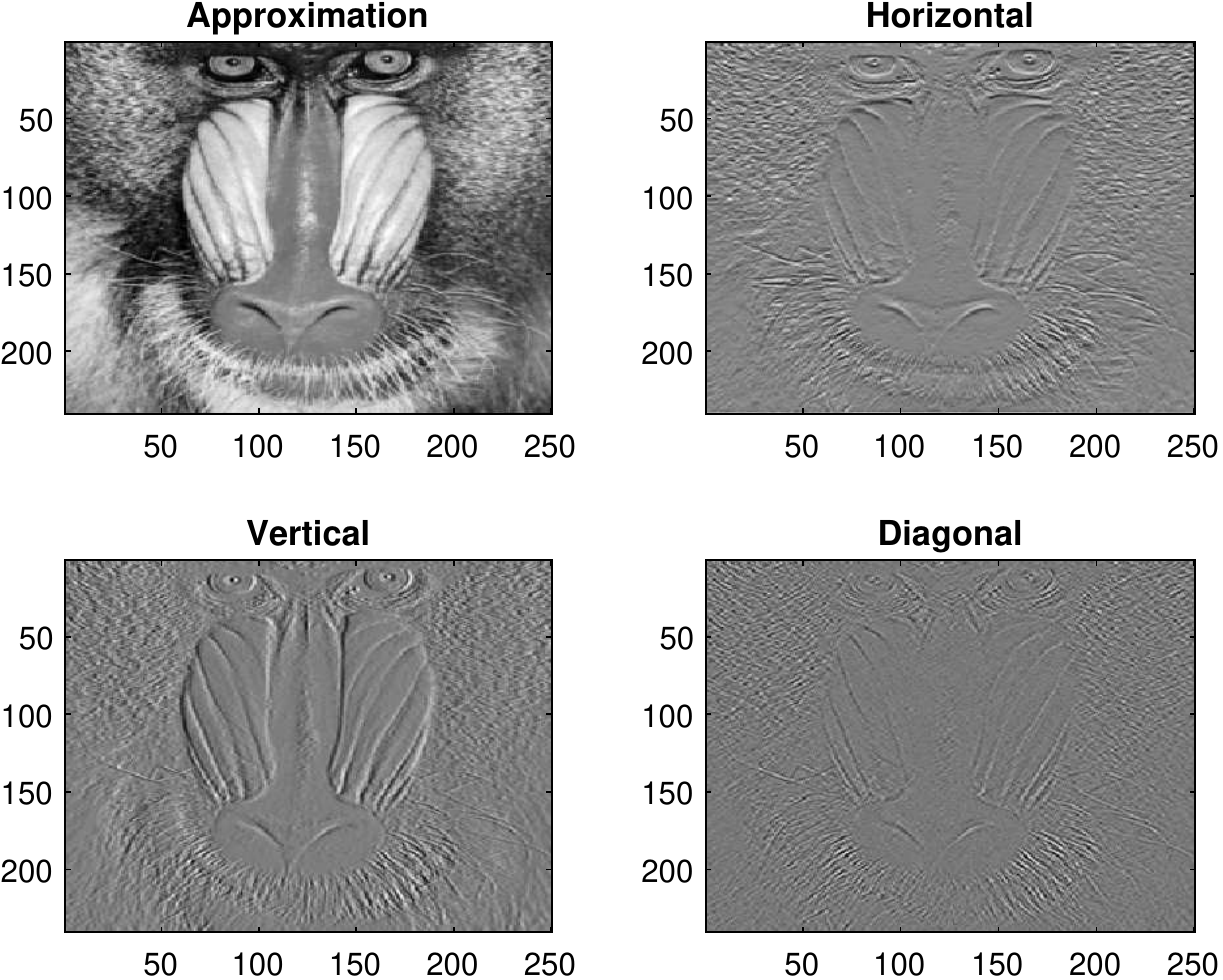}
		\caption{Haar}
	\end{subfigure}
	\begin{subfigure}{0.4\columnwidth}
		\includegraphics[width=\columnwidth]{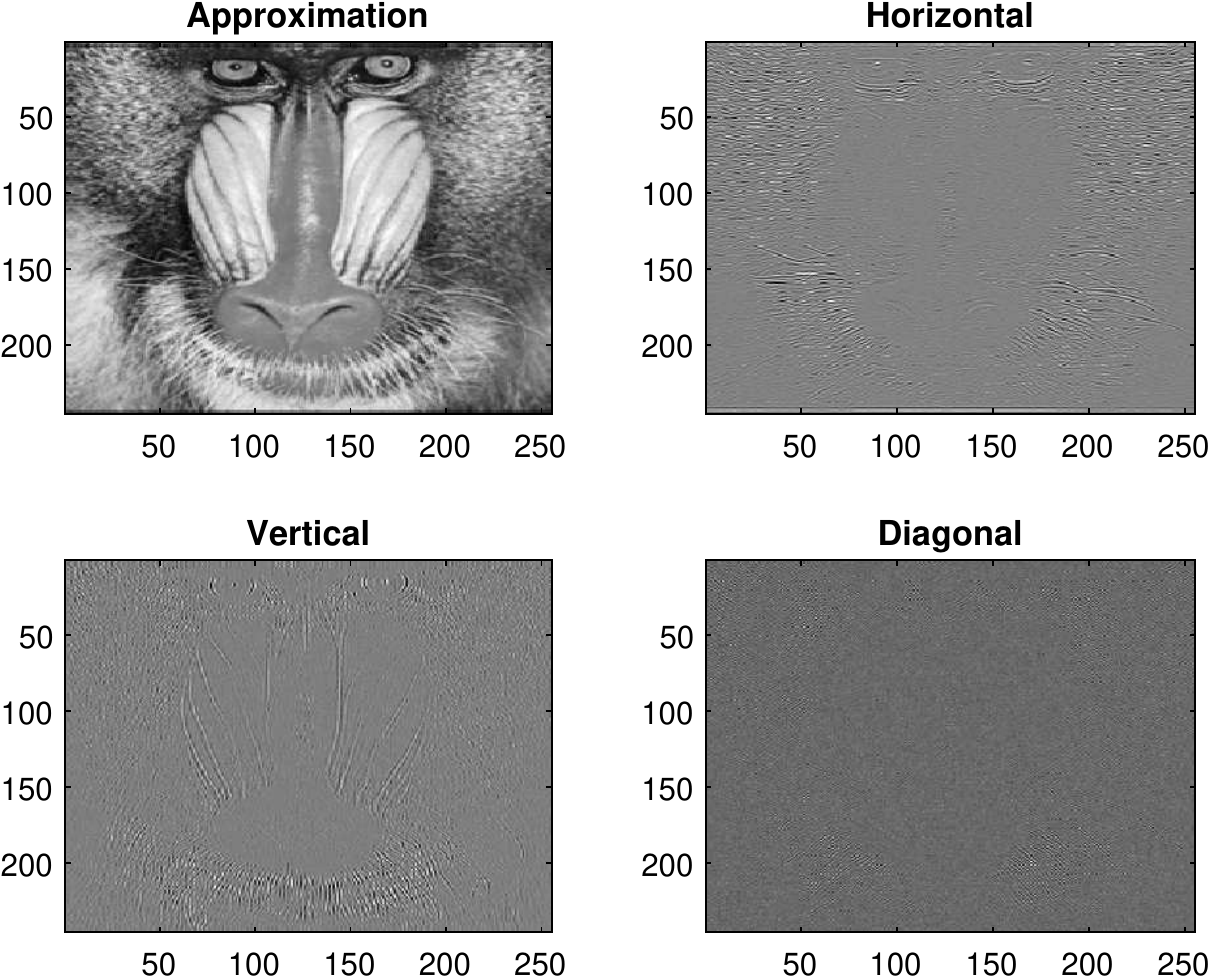}
		\caption{Coiflet 2}
	\end{subfigure}
	\begin{subfigure}{0.4\columnwidth}
		\includegraphics[width=\columnwidth]{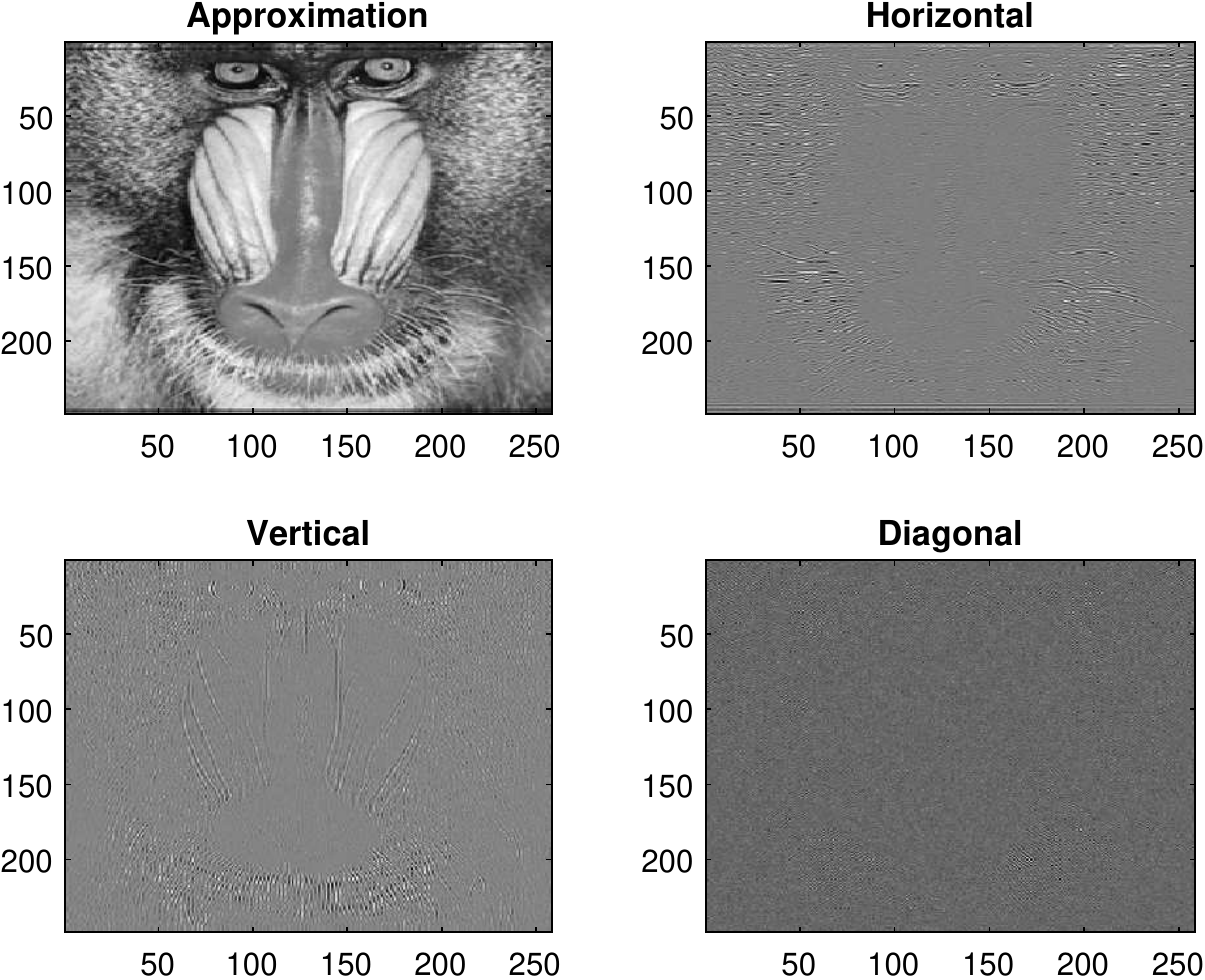}
		\caption{Coiflet 3}
	\end{subfigure}
	\begin{subfigure}{0.4\columnwidth}
		\includegraphics[width=\columnwidth]{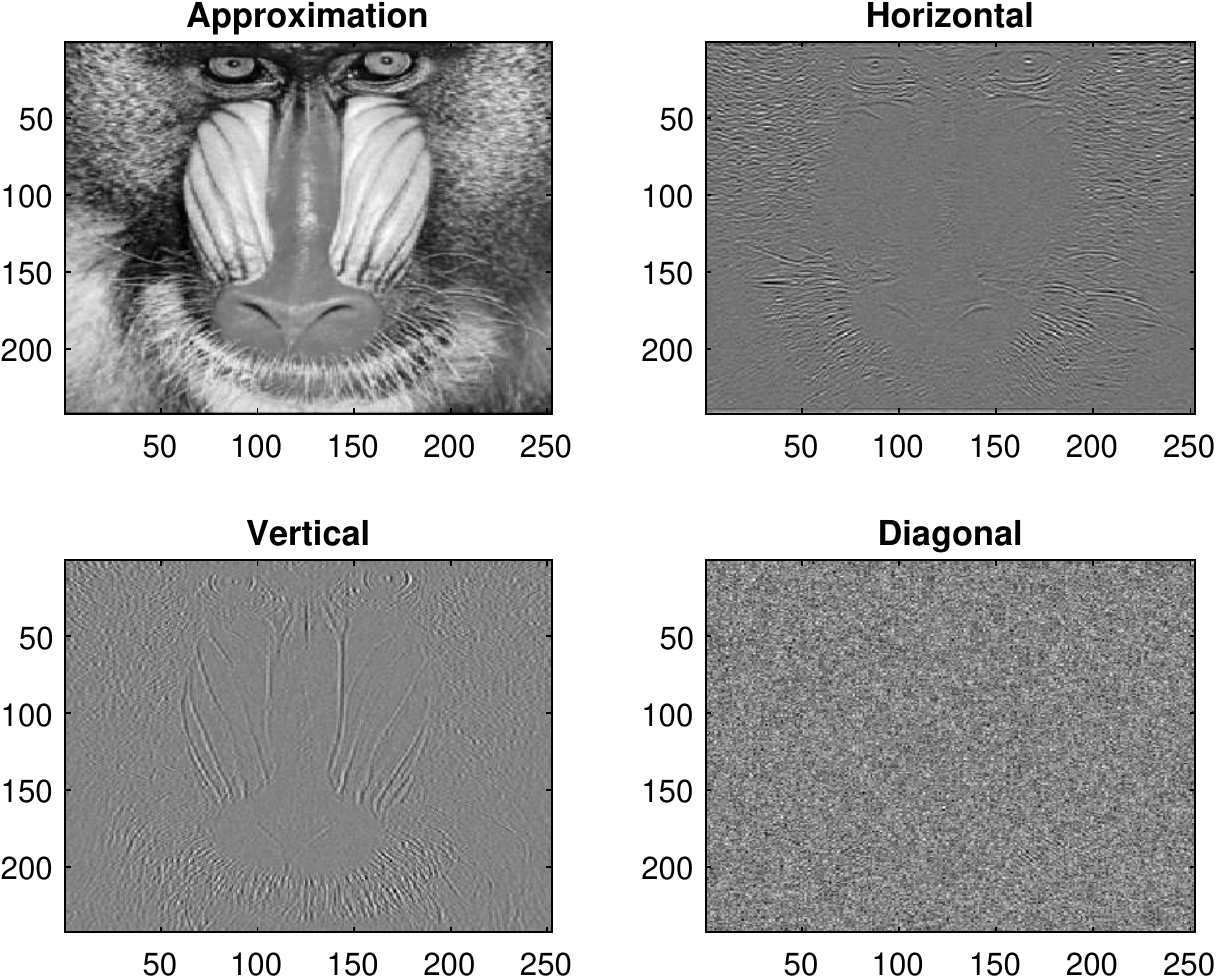}
		\caption{Bior 1.3}
	\end{subfigure}
	\caption{Four exemplary wavelets and their decomposition.}
	\label{fig:waveletDecGraphs}
\end{figure}

Wavelets are useful tools for conducting frequency and scale analysis, and make excellent tools for removing noise from images. A wavelet forms an orthogonal basis for approximating a signal. The continuous wavelet transform with scale parameter $\alpha$ and translation parameter $\beta$ is given by Eq. \ref{eq:CWT}
\begin{equation}
\label{eq:CWT}
\centering
G(\alpha, \beta) = \int x(t)\psi_{\alpha,\beta}^{*}(t)dt
\end{equation}
and the inverse of this operation to reconstruct the signal is given by Eq. \ref{eq:iCWT}.
\begin{equation}
\label{eq:iCWT}
\centering
x(t) = \iint G(\alpha, \beta)\psi_{\alpha,\beta}(t)d\alpha d\beta
\end{equation}
Given a wavelet $\psi_{\alpha,\beta}(n)$ with scale and translation parameters $\alpha$ and $\beta$ and a signal, $x_1, x_2, \ldots, x_N$ of length $N$, then the discrete wavelet transform (DWT) is given by Eq. \ref{eq:wvApprox}
\begin{equation}
\label{eq:wvApprox}
\centering
F(\alpha, \beta)=\sum_{i=1}^{N}x_i\psi_{\alpha,\beta}(i)
\end{equation}
with $\psi_{\alpha,\beta}(i)$ computed as Eq. \ref{eq:wavelet}
\begin{equation}
\label{eq:wavelet}
\centering
\psi_{\alpha,\beta}(i) = 2^{-\alpha/2}\psi(2^{-\alpha}i-\beta).
\end{equation}
The discrete wavelet deconstruction is computed using the filter bank paradigm, wherein the analyzing wavelet is viewed as a band-pass filter \cite{valens1999really}. By convoluting the wavelet filters over the image, the wavelet decomposition can be computed simply. For example, the $2 \times 2$ matrix representing the Haar wavelet is given in Eq. \ref{eq:HaarFilter}.
\begin{equation}
\label{eq:HaarFilter}
\centering
\begin{bmatrix}
1 & 1 \\
1 & -1
\end{bmatrix}
\end{equation}
This results in two signals of half length with are the approximation signal (low band) and the detail signal (high band). A 2-D wavelet transform can be computed by treating the 2-D signal as a sequence of 1-D row and column signals. First, the rows are transformed by a 1-D wavelet transform and then the columns. This creates four signals, one approximation signal (low-low band) and three detail signals (low-high, high-low, and high-high bands). The flow diagrams depicting the filter bank approach for wavelet decomposition and reconstruction are given in Figures \ref{fig:dec} and \ref{fig:rec}.
Examples of wavelet decomposition with some different analyzing wavelets are given in Figure \ref{fig:waveletDecGraphs}.

\subsection{GHM Multi-Level Wavelet}
\label{multiwaevlets}
Single-level wavelets discussed in Section \ref{singlewavelets} consist of one scaling function and one wavelet function. This can limit the resolution of the decomposition because the wavelet function has fixed support length. In contrast, multi-level wavelets can have more than one scaling function and wavelet function. This allows for a wavelet decomposition that can blend small and large support vectors. A popular wavelet for multi-level wavelet decompositions is the GHM wavelet \cite{ben2015high, wang2005multispectral,yajnik2015numerical, deivalakshmi2016undecimated}.

The GHM wavelet function and scaling function are given in Eq. \ref{ghmeq}. The filter banks $H_k$ and $G_k$, where $k=2$ is the number of scaling and wavelet functions are given in Eq. \ref{GHMfilters}\cite{deivalakshmi2016undecimated}.

\begin{align}
        \phi (t) &= \sqrt{2}\sum_{-\inf}^{\inf} H_k\phi(2t - k) \\
    \psi (t) &= \sqrt{2}\sum_{-\inf}^{\inf} G_k\phi(2t - k)
    \label{ghmeq}
\end{align}

\begin{align}
    H_k &= \begin{bmatrix}
    h_0(2k) & h_0(2k+1) & h_0(2k+2) & h_0(2k+3)\\
    h_1(2k) & h_1(2k+1) & h_0(2k+2) & h_0(2k+3)
    \end{bmatrix} \\
    G_k &= \begin{bmatrix}
    g_0(2k) & g_0(2k+1) & g_0(2k+2) & g_0(2k+3)\\
    g_1(2k) & g_1(2k+1) & g_1(2k+2) & g_1(2k+3)
    \end{bmatrix}
    \label{GHMfilters}
\end{align}

\begin{figure}
    \centering
    \includegraphics[width=0.3\columnwidth]{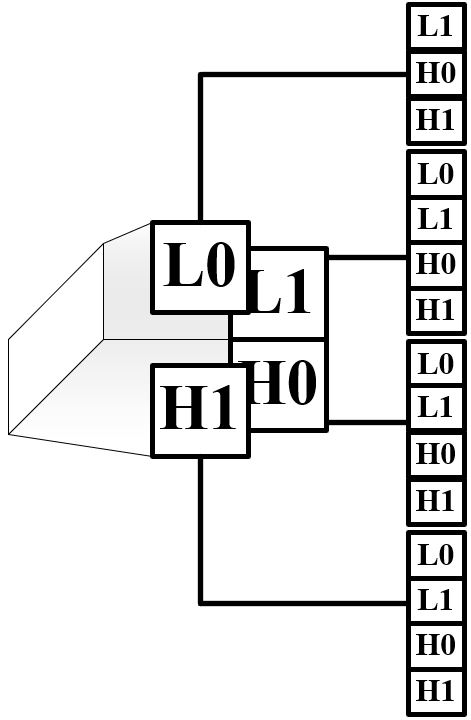}
    \caption{Two level image decomposition with the GHM wavelet.}
    \label{fig:GHMwavelet}
\end{figure}

The kernel matrices for the GHM multi-level wavelet are $H_1, H_2$ and $G_1, G_2$ given in Eq. \ref{GHMfilterbank}. The kernels can be convolved with the input image in a filter bank approach to perform the wavelet decomposition. The two stage filter bank approach is given in Figure \ref{fig:GHMwavelet}.

\begin{align}
    H_1 &= \begin{bmatrix}
    \frac{3}{5\sqrt{2}} & \frac{4}{5} & \frac{3}{5\sqrt{2}} & 0 \\
    \frac{-1}{20} & \frac{-3}{10\sqrt{2}} & \frac{9}{20} & \frac{1}{\sqrt{2}} \\
    \end{bmatrix} \\
    G_1 &= \begin{bmatrix}
    \frac{-1}{20} & \frac{-3}{10\sqrt{2}} & \frac{9}{20} & \frac{-1}{\sqrt{2}} \\
    \frac{1}{10\sqrt{2}} & \frac{3}{10} & \frac{-9}{10\sqrt{2}} & 0
    \end{bmatrix} \\
    H_2 &= \begin{bmatrix}
    0 & 0 & 0 & 0 \\
    \frac{9}{20} & \frac{-3}{10\sqrt{2}} & \frac{-1}{20} & 0
    \end{bmatrix} \\
    G_2 &= \begin{bmatrix}
    \frac{9}{20} & \frac{-3}{10\sqrt{2}} & \frac{-1}{20} & 0 \\
    \frac{9}{10\sqrt{2}} & \frac{-3}{10} & \frac{-1}{10\sqrt{2}} & 0
    \end{bmatrix}
    \label{GHMfilterbank}
\end{align}
\begin{figure*}
	\centering
	\includegraphics[width=\linewidth]{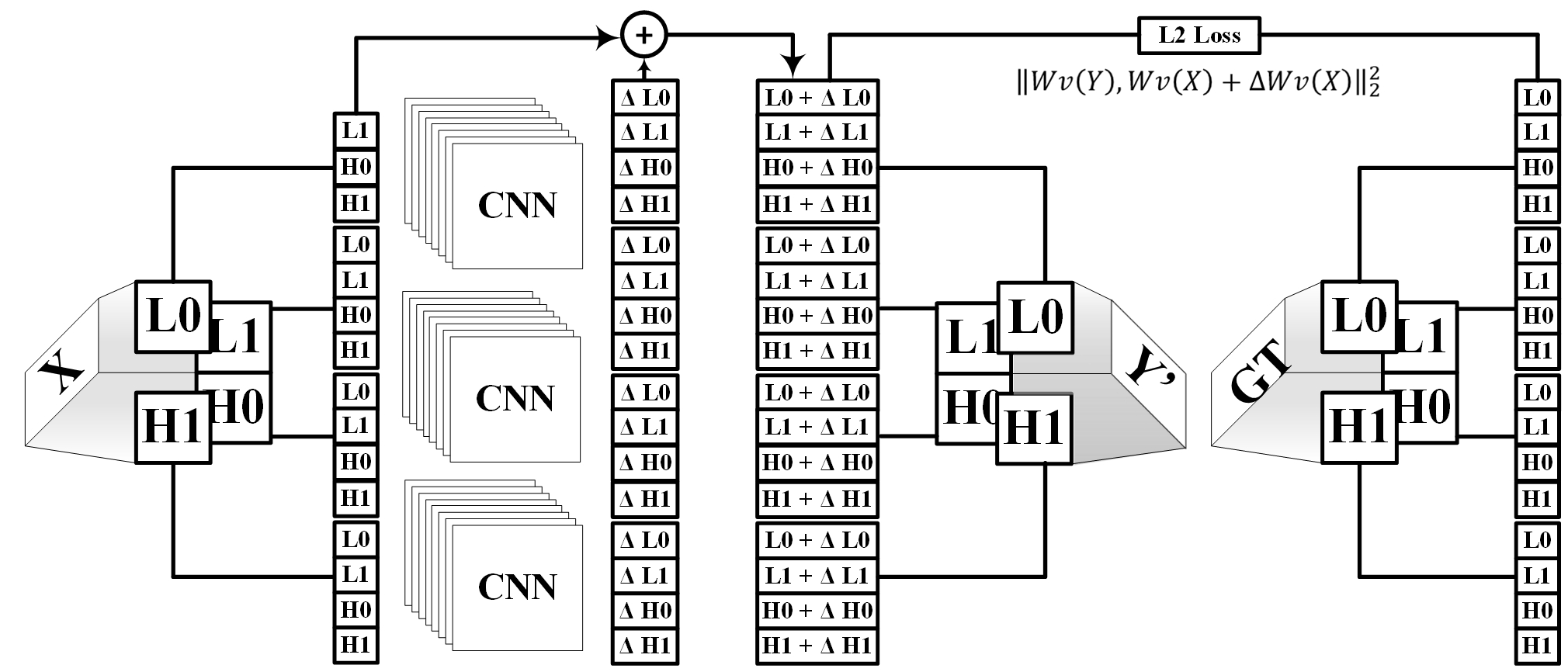}
	\caption{GHM Multi-Wavelet Network Architecture}
	\label{fig:GHMnetwork}
\end{figure*}

\subsection{Objective Error Measures}
The seven objective error measures used in this analysis are peak signal to noise ratio (PSNR), structural similarity (SSIM), gradient similarity (GSM), feature similarity (FSIM), spectral similarity (SR-SIM), most apparent distortion (MAD), and visual information fidelity (VIF). They are described mathematically here. The PSNR is the most widely used measure of image quality because it is simple to calculate and provides a metric for quantitative comparison between super resolution techniques \cite{vansurvey}. PSNR is based on the mean squared error. If $x_{jk}$ is the $j^\textit{th}$ pixel in the $k^\textit{th}$ channel of the ground truth image and $\hat{x}_{jk}$ is the corresponding pixel and channel of the approximated image in images of $p$ pixels and $c$ channels then the mean squared error is given by Eq. \ref{eq:MSE}.
\begin{equation}
\label{eq:MSE}
\centering
MSE = \frac{1}{pc}\sum_{j=1}^{p}\sum_{k=1}^{c}(x_{jk}-\hat{x}_{jk})^2.
\end{equation}
Then, the PSNR can be determined from the mean square error, $MSE$, and the maximum signal value, $m$ as Eq. \ref{eq:PSNR}
\begin{equation}
\label{eq:PSNR}
\centering
PSNR=10\log_{10}(\frac{m^2}{MSE}).
\end{equation}
Wang et al. notes that PSNR does not give a good metric for perceived quality of an image \cite{wang2004image}. Therefore it is not sufficient to draw image quality conclusions simply from PSNR. Wang et al. proposes the measure of structural similarity (SSIM) which constructs the mean of the similarity between patches of the compared images \cite{wang2004image}. The SSIM can be given using the mean, variance, and correlation of the window $w$ of images $x$ and $\hat{x}$ denoted $\mu_x$, $\sigma_x$, and $\sigma_{x\hat{x}}$, respectively. Then, the SSIM is given by Eq. \ref{eq:SSIM}
\begin{equation}
\label{eq:SSIM}
\centering
SSIM(x, \hat{x}, w) = \frac{(2\mu_{x_w}\mu_{\hat{x}_w}+\alpha)(2\sigma_{x_w\hat{x}_w}+\beta)}{(\mu_{x_w}^2+\mu_{\hat{x}_w}^2+\alpha)(\sigma_{x_w}^2+\sigma_{\hat{x}_w}^2+\beta)}
\end{equation}
where $\alpha$ and $\beta$ are constants chosen to be small to avoid zero in the denominator.
A mean SSIM can be generated by computing the mean SSIM of each window. Given the set of windows $W$, then the mean SSIM or MSSIM is given by Eq. \ref{eq:MSSIM}
\begin{equation}
\label{eq:MSSIM}
\centering
MSSIM(x, \hat{x}) = \frac{1}{|W|}\sum_{w\in W}SSIM(x, \hat{x}, w)
\end{equation}
Liu, Lin, and Narwaria proposed an image quality assessment using gradient similarity, GSM. \cite{liu2011image} uses four $5\times 5$ kernels for calculating the gradient value. The kernels are given in Eq. \ref{eq:5x5kernels}.
\begin{equation}
\label{eq:5x5kernels}
\centering
\begin{aligned}
\begin{bmatrix}
0 & 0 & 0 & 0 & 0 \\
1 & 3 & 8 & 3 & 1 \\
0 & 0 & 0 & 0 & 0 \\
-1 & -3 & -8 & -3 & -1 \\
0 & 0 & 0 & 0 & 0 
\end{bmatrix}&
\begin{bmatrix}
0 & 0 & 1 & 0 & 0 \\
0 & 8 & 3 & 0 & 0 \\
1 & 3 & 0 & -3 & -1 \\
0 & 0 & -3 & -8 & 0 \\
0 & 0 & -1 & 0 & 0 \\
\end{bmatrix}
\end{aligned}
\centering
\begin{aligned}
\begin{bmatrix}
0 & 0 & 1 & 0 & 0 \\
0 & 0 & 3 & 8 & 0 \\
-1 & -3 & 0 & 3 & 1 \\
0 & -8 & -3 & 0 & 0 \\
0 & 0 & -1 & 0 & 0 
\end{bmatrix}&
\begin{bmatrix}
0 & 1 & 0 & -1 & 0 \\
0 & 3 & 0 & -3 & 0 \\
0 & 8 & 0 & -8 & 0 \\
0 & 3 & 0 & -3 & 0 \\
0 & 1 & 0 & -1 & 0 
\end{bmatrix}
\end{aligned}
\end{equation}
Then, with $mmean$ as the mean of a matrix, the gradient value on the set of kernels $K$ for rows, and similarly columns, is given by Eq. \ref{eq:gradientValue}
\begin{equation}
\label{eq:gradientValue}
\centering
g_r = \max_{k \in K}\{mmean(|r \cdot k)|\}
\end{equation}
With $D = \frac{|g_r - g_c|}{\max\{g_r, g_y\}}$ and near-zero constant $C$, the gradient similarity metric in \cite{liu2011image} is given in Eq. \ref{eq:GSM}.
\begin{equation}
\label{eq:GSM}
\centering
g(r, c) = \frac{2(1-D)+C}{1+(1-D)^2 + C}
\end{equation}
The feature similarity metric, FSIM, was proposed by Zhang \cite{zhang2011fsim}. FSIM computes a weighted average of local structural similarity scores as opposed to SSIM which gives equal weight to each local score. The local structural similarity scores are derived from two features: phase congruency (PC) and gradient magnitude (GM) \cite{zhang2011fsim}. A subset of the image space, or window, with a high phase congruency is likely to contain high impact features so it is used as a weight for the FSIM metric. Given the set of local structure windows $W$, and the similarity metrics $S_{PC}$ and $S_{GM}$ then the feature similarity is given by Eq. \ref{eq:FSIM}
\begin{equation}
\label{eq:FSIM}
\centering
FSIM = \frac{\sum_{w \in W}S_{PC}(w)S_{GM}(w)\cdot \max_{i \in \{x, \hat{x}\}}\{PC_i(w)\}}{\sum_{w \in W}\max_{i \in \{x, \hat{x}\}}\{PC_i(w)\}}
\end{equation}
Spectral residual similarity (SR-SIM), developed by Zhang and Li \cite{zhang2012sr} is also used to measure the performance of the wavelets. SR-SIM uses spectral residual visual saliency (SRVS) and gradient modulus as features for the image quality assessment. Zhang and Li adopt the Scharr gradient operator with gradient modulus $G(\boldsymbol{x}) = \sqrt{G_x^2(\boldsymbol{x})+G_y^2(\boldsymbol{x})}$. The use of SRVS is justified due to the relationship between an image's visual saliency map and its perceived visual quality \cite{zhang2012sr}.

The most apparent distortion or MAD metric \cite{larson2010most} attempts to merge the qualities of PSNR and SSIM strategies. Larson et al. combines a technique for identifying distortions in high quality images and a technique for low quality images. In the former, local MSE approach is used on the perceived luminance to construct a visible distortions map. The map is concentrated into a scalar metric using Euclidean distance, $d_{lum}$. In the latter, log-Gabor filters are used to decompose the compared images and the differences in the filter responses are used to generate a quality metric, $d_{gabor}$ \cite{larson2010most}. The two strategies are combined with a blending parameter $\gamma$ by Eq. \ref{eq:MAD}.
\begin{equation}
\label{eq:MAD}
\centering
MAD = d_{lum}^\gamma d_{gabor}^{1-\gamma}
\end{equation}

The visual information fidelity (VIF) measure views image quality assessment as a measure of image information. Specifically, VIF quantifies the amount of information contained in the ground truth image and then measures how much of the ground truth information is present in the reconstructed image \cite{sheikh2006image}. 
\subsection{Network Architecture and Hyperparameters}
\begin{figure}
	\centering
	\includegraphics[width=\columnwidth]{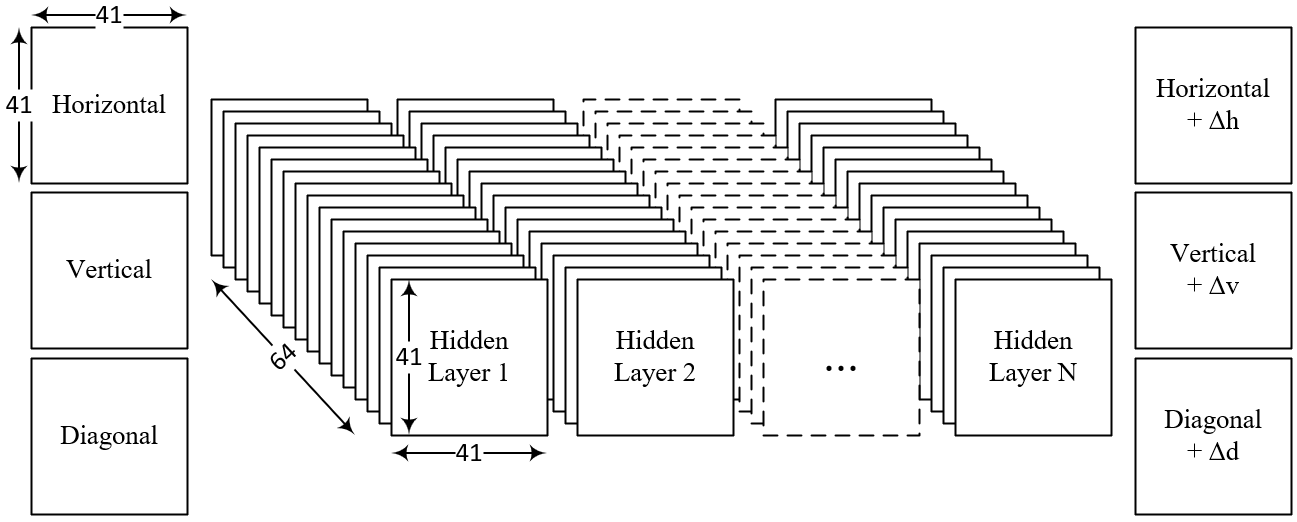}
	\caption{Wavelet-based Convolutional Neural Network Architecture.}
	\label{fig:networkarchitecture}
\end{figure}

The network architecture is given in Figure \ref{fig:networkarchitecture}. The network architecture for the GHM multi-wavelet approach is given in Figure \ref{fig:GHMnetwork}. There are 10 convolutional layers which learn the change from the three low resolution wavelet detail bands to the corresponding high resolution detail bands. Each convolutional layer consists of 64 filters with kernel size $3\times 3$. Following each convolutional layer is a ReLU activation function. The equation representing the forward pass of the network through layer $l$ is given in Eq. \ref{eq:forwardpass}.

\begin{equation}
    y_l = \max(0, x_l \ast K_l)
    \label{eq:forwardpass}
\end{equation}

The approximate high resolution wavelet decomposition is given by adding the network output to the detail bands from the low resolution input as shown in Figure \ref{fig:GHMnetwork}. The network is trained with 50 epochs using the Adam optimizer with learning rate $0.01$ and moment decay $\beta=0.9$. The loss function used is the L2-norm between the predicted wavelet coefficients and the high-resolution ground truth wavelet coefficients in Eq. \ref{eq:L2Loss} where $Wv(\cdot)$ is the wavelet decomposition and $\Delta Wv(x)$ is the forward prediction of the network. 
\begin{equation}
    \mathcal{L}_{L2}(y,\hat{y}) = || Wv(y), Wv(x) + \Delta Wv(x) ||_{2}^{2}
    \label{eq:L2Loss}
\end{equation}
\section{Experimental Evaluation}
\label{sec:experiment}
This section contains the results of the experiments described in \ref{sec:methods}. The results are presented such that a comparison between each wavelet's performance can be easily seen. We consider seven different objective error measures chosen from the literature. For a description of each, the interested reader may refer to Section \ref{sec:methods} or to their respective papers.
\subsection{Single-Level Wavelet Experiments}
Many single-level wavelets exist, however only the Haar wavelet is used in the deep image super resolution approach by Guo et al. \cite{dwsr}. This section presents mean scores for image quality assessments on the Set14 data set using a variety of different wavelets. The wavelets tested are from Haar, Daubechies, Biorthogonal, Reerse Biorthogonal, Coiflets, and Symlets families.

While the decomposed images on different wavelets vary greatly as shown in Figure \ref{fig:waveletDecGraphs}, the results show that the single-level wavelets perform similarly. This indicates that the structure of the input representation is more important than the properties of the embedding. Haar is a commonly used wavelet for image signal processing in part because the decomposition preserves the edges of the input image clearly in the decomposition. However the reverse biorthogonal wavelet 1.3 in Figure \ref{fig:waveletDecGraphs} does not have clear edges in the decomposed images but still achieves similar results to the Haar wavelet.

 The network was trained on the DIV2K dataset \cite{agustsson2017ntire} and tested on the Set14 test set, a standard test set for super resolution. Each predicted high resolution image is compared with the high resolution ground truth with several image quality assessments detailed in Section \ref{sec:methods}. The results were compared to a re-implementation of the Haar wavelet DWSR \cite{dwsr} and a bicubic interpolation with no neural network as a baseline for performance. The mean score results from Set14 reconstructions are given in Table \ref{tab:singlelevelwv} in Appendix A. 

\subsection{GHM Experiments}
The GHM wavelet was used as preprocessing for the convolutional network as given in Section \ref{sec:methods}. The network was trained on the DIV2K dataset \cite{agustsson2017ntire} and tested on the Set14 test set, a standard test set for super resolution. Each predicted high resolution image is compared with the high resolution ground truth with several image quality assessments detailed in Section \ref{sec:methods}. The results were compared to a re-implementation of the Haar wavelet DWSR \cite{dwsr} and a bicubic interpolation with no neural network as a baseline for performance. The mean score results from Set14 reconstructions are given in Table \ref{tab:GHMExp}. The proposed GHM multi-level wavelet network outperforms the baseline and Haar wavelet approach on the Set14 data set.
\begin{table}[h]
    \centering
    \begin{tabular}{|c|c|c|c|}
    \hline
         IQA & Bicubic & Haar* \cite{dwsr} & GHM (Proposed) \\
         \hline
         \hline
         PSNR & 25.5616 & \textbf{26.5901} & 26.4518 \\ 
         \hline
         SSIM \cite{wang2004image} & 0.8130 & \textbf{0.8560} & 0.7917 \\
         \hline
         FSIM \cite{zhang2011fsim} & 0.9418 & 0.9616 & \textbf{0.9868} \\
         \hline
         GSM \cite{liu2011image} & 0.9917 & 0.9945 & \textbf{0.9978} \\
         \hline
         MAD \cite{larson2010most} & 0.3962 & 0.4160 & \textbf{0.4183} \\
         \hline
         SRSIM \cite{zhang2012sr} & 0.9701 & 0.9807 & \textbf{0.9946} \\
         \hline
         VIF \cite{sheikh2006image} & 0.5210 & 0.6243 & \textbf{0.6554} \\
         \hline
    \end{tabular}
    \caption{The proposed GHM multi-level wavelet network outperforms the Haar and baseline in 6 image quality assessments on the Set14 data set. * is a re-implementation of \cite{dwsr}.}
    \label{tab:GHMExp}
\end{table}
\section{Conclusion}
\label{sec:conclusion}
In this paper, a GHM multi-level wavelet is applied to deep image super resolution. The proposed approach outperforms the single-level wavelets including the Haar wavelet when tested on the Set14 data set. An experimental evaluation was performed for 37 single-level wavelets and the GHM multi-level wavelet trained on the DIV2K dataset. Seven image quality assessments were used to compare the high resolution image reconstruction. The GHM wavelet outperformed the Haar wavelet and the baseline bicubic interpolation in feature similarity \cite{zhang2011fsim}, gradient similarity \cite{liu2011image}, most apparent distortion \cite{larson2010most}, spectral similarity \cite{zhang2012sr}, and visual information fidelity \cite{sheikh2006image}. The single-level wavelet performed similarly to the Haar wavelet which indicates that the structure of the wavelet input is more important for the network to learn to reconstruct the high resolution image rather than properties such as edge preservation in the embeddings. The increased quality from using the GHM multi-wavelet may be because it produces 16 subbands rather than 4 subbands in a single-level wavelet. Extending the experiment to other test and training sets remains as future work.
\bibliographystyle{unsrt}  
\bibliography{references} 
\clearpage
\appendix
\section{Appendix A}
\label{sec:singlewvresults}
\subsection{Single-Level Wavelet Results}

The results of the single-level wavelet testing are presented here.

\begin{table}[!h]
    \centering
    \begin{tabular}{|c|c|c|c|c|c|c|c|}
    \hline
         & PSNR & SSIM & FSIM & GSM & MAD & SRSIM & VIF \\
         \hline
         \hline
Bicubic & 
25.5616 &
0.8130 &
0.9418 &
0.9917 &
0.3962 &
0.9701 &
0.5210 \\ \hline
Haar &
26.5901 &
0.8560 &
0.9616 &
0.9945 &
0.4160 &
0.9807 &
0.6243 \\ \hline
bior2.6 &
26.6149 &
0.85692 &
0.96182 &
0.99455 &
0.41761 &
0.98085 &
0.62595 \\ \hline
coif1 &
26.5599 &
0.85573 &
0.96138 &
0.99451 &
0.41646 &
0.98066 &
0.6221 \\ \hline
coif2 &
26.5848 &
0.85594 &
0.96151 &
0.99451 &
0.41651 &
0.98069 &
0.62269 \\ \hline
coif3 &
26.5959 &
0.85583 &
0.96131 &
0.9945 &
0.41624 &
0.98063 &
0.62248 \\ \hline
coif4 &
26.5878 &
0.8554 &
0.96111 &
0.99445 &
0.4172 &
0.98048 &
0.62204 \\ \hline
coif5 &
26.592 &
0.85577 &
0.96126 &
0.99452 &
0.41688 &
0.98062 &
0.62296 \\ \hline
db10 &
26.5856 &
0.85574 &
0.96121 &
0.9945 &
0.41674 &
0.98057 &
0.62341 \\ \hline
db18 &
26.4508 &
0.85074 &
0.95945 &
0.99426 &
0.41377 &
0.97962 &
0.61268 \\ \hline
db19 &
26.4274 &
0.8498 &
0.9592 &
0.99424 &
0.41342 &
0.97944 &
0.61179 \\ \hline
db1 &
26.5901 &
0.85603 &
0.96156 &
0.9945 &
0.41603 &
0.98075 &
0.62429 \\ \hline
db20 &
26.4459 &
0.85054 &
0.95907 &
0.99425 &
0.41344 &
0.97948 &
0.61185 \\ \hline
db2 &
26.5503 &
0.85509 &
0.96102 &
0.99446 &
0.41601 &
0.98046 &
0.62089 \\ \hline
db3 &
26.6063 &
0.85656 &
0.96157 &
0.99453 &
0.41761 &
0.98078 &
0.6255 \\ \hline
db4 &
26.5996 &
0.85602 &
0.96142 &
0.99448 &
0.41693 &
0.98065 &
0.62417 \\ \hline
db5 &
26.606 &
0.85606 &
0.96145 &
0.99451 &
0.41687 &
0.98071 &
0.6232 \\ \hline
db6 &
26.619 &
0.85726 &
0.96192 &
0.99459 &
0.41772 &
0.98092 &
0.62548 \\ \hline
db7 &
26.589 &
0.85656 &
0.96158 &
0.99453 &
0.41579 &
0.98077 &
0.62621 \\ \hline
db8 &
26.5879 &
0.85585 &
0.96132 &
0.99448 &
0.4172 &
0.98059 &
0.62416 \\ \hline
rbio2.6 &
26.6017 &
0.85577 &
0.96132 &
0.99447 &
0.41686 &
0.98062 &
0.62114 \\ \hline
rbio2.8 &
26.6126 &
0.85681 &
0.96176 &
0.99455 &
0.4174 &
0.98083 &
0.62408 \\ \hline
rbio3.1 &
26.5542 &
0.85587 &
0.96155 &
0.99449 &
0.41732 &
0.98067 &
0.62351 \\ \hline
rbio3.3 &
26.5674 &
0.85453 &
0.96053 &
0.9944 &
0.41557 &
0.98025 &
0.61899 \\ \hline
rbio3.5 &
26.5607 &
0.85401 &
0.96066 &
0.99438 &
0.41623 &
0.98021 &
0.61745 \\ \hline
sym12 &
26.6098 &
0.85681 &
0.96175 &
0.99455 &
0.41697 &
0.98084 &
0.62486 \\ \hline
sym13 &
26.6173 &
0.85693 &
0.96197 &
0.99457 &
0.41718 &
0.981 &
0.62468 \\ \hline
sym14 &
26.5957 &
0.85599 &
0.96134 &
0.99448 &
0.41688 &
0.9806 &
0.62299 \\ \hline
sym15 &
26.6065 &
0.85615 &
0.96128 &
0.99449 &
0.41737 &
0.9806 &
0.62411 \\ \hline
sym2 &
26.5927 &
0.85659 &
0.96165 &
0.99456 &
0.417 &
0.98078 &
0.626 \\ \hline
sym3 &
26.6149 &
0.85675 &
0.9615 &
0.99453 &
0.4179 &
0.98074 &
0.62558 \\ \hline
sym4 &
26.5753 &
0.85536 &
0.96131 &
0.99448 &
0.41627 &
0.98062 &
0.62153 \\ \hline
sym5 &
26.5569 &
0.85459 &
0.96077 &
0.99441 &
0.41623 &
0.98035 &
0.62184 \\ \hline
sym6 &
26.616 &
0.85648 &
0.96157 &
0.99454 &
0.41722 &
0.98079 &
0.6246 \\ \hline
sym7 &
26.6111 &
0.85665 &
0.96182 &
0.99455 &
0.41803 &
0.98087 &
0.62701 \\ \hline
sym8 &
26.5989 &
0.85604 &
0.96152 &
0.99451 &
0.41709 &
0.98071 &
0.62364 \\ \hline
sym9 &
26.5916 &
0.85586 &
0.96132 &
0.9945 &
0.41622 &
0.98067 &
0.62267 \\ \hline

    \end{tabular}
    \caption{The single level wavelets perform very similarly.}
    \label{tab:singlelevelwv}
\end{table}

\end{document}